\documentclass[prb,twocolumn,a4paper]{revtex4}
\usepackage{amssymb}
\usepackage{amsmath}
\usepackage[pdftex]{graphicx}

\providecommand{\unit}[1]{\,\mbox{#1}} 
\providecommand{\ket}[1]{\vert #1\rangle} 
\providecommand{\bra}[1]{\langle #1\vert} 
\providecommand{\braket}[2]{\langle #1\vert #2 \rangle} 

\begin{document}

\title{\vspace*{-0.9cm} Collapse of $\rho_{xx}$ ringlike structures in 2DEGs under tilted magnetic fields}
\author{Gerson J.~Ferreira}
\author{J.~Carlos Egues}
\affiliation{Departamento de F\'{\i}sica e Inform\'{a}tica, Instituto de F\'{\i}sica de S\~{a}o Carlos, Universidade de S\~{a}o Paulo, 13560-970 S\~{a}o Carlos, S\~{a}o Paulo, Brazil}

\begin{abstract}
In the quantum Hall regime, the longitudinal resistivity $\rho_{xx}$ plotted as a density--magnetic-field ($n_{2D}-B$) diagram displays ringlike structures due to the crossings of two sets of spin split Landau levels from different subbands [e.g., Zhang \textit{et al.}, Phys. Rev. Lett. \textbf{95}, 216801 (2005)]. For tilted magnetic fields, some of these ringlike structures ``shrink'' as the tilt angle is increased and fully collapse at $\theta_c \approx 6^\circ$. Here we theoretically investigate the topology of these structures via a non-interacting model for the 2DEG. We account for the inter Landau-level coupling induced by the tilted magnetic field via perturbation theory. This coupling results in anti-crossings of Landau levels with parallel spins. With the new energy spectrum, we calculate the corresponding $n_{2D}-B$ diagram of the density of states (DOS) near the Fermi level. We argue that the DOS displays the same topology as $\rho_{xx}$ in the $n_{2D}-B$ diagram. For the ring with filling factor $\nu=4$, we find that the anti-crossings make it shrink for increasing tilt angles and collapse at a large enough angle. Using effective parameters to fit the $\theta = 0^\circ$ data, we find a collapsing angle $\theta_c \approx 3.6^\circ$. Despite this factor-of-two discrepancy with the experimental data, our model captures the essential mechanism underlying the ring collapse.
\end{abstract}

\preprint{Journal reference: preprint}
\maketitle
In the quantum Hall regime with tilted magnetic fields making an angle $\theta$ with the normal of the two-dimensional electron gas (2DEG), the transverse magnetic field $B_{\bot} = B \cos\theta$ quantizes the planar degrees of freedom into macroscopically degenerate Landau levels (LLs) \cite{Sarma97,Bastard88}. Whenever LLs cross near the Fermi level an interplay between Zeeman, Coulomb and temperature energy scales may give rise to quantum Hall ferromagnetic phase transitions \cite{Jungwirth00, Freire07PRL}. Recently, systems displaying these transitions have been studied experimentally \cite{Poortere00, Jaroszynski02, Ritchie96, Muraki01, Zhang05, Ellenberger06, Zhang06_PRB, Zhang06_PRL, ZhangSU4-07,Muraki07} and theoretically \cite{Jungwirth00,Jungwirth01,Gerson06,Freire07PRL}. LL crossings can be obtained by different methods, e.g.: (i) by tilting the magnetic field with respect to the 2DEG \cite{Poortere00} and (ii) by varying the electron density in order to make the second subband occupied \cite{Ritchie96, Muraki01, Zhang05, Ellenberger06}.

\begin{figure}[ht!]
   \centering
   \includegraphics[width=8cm,keepaspectratio=true]{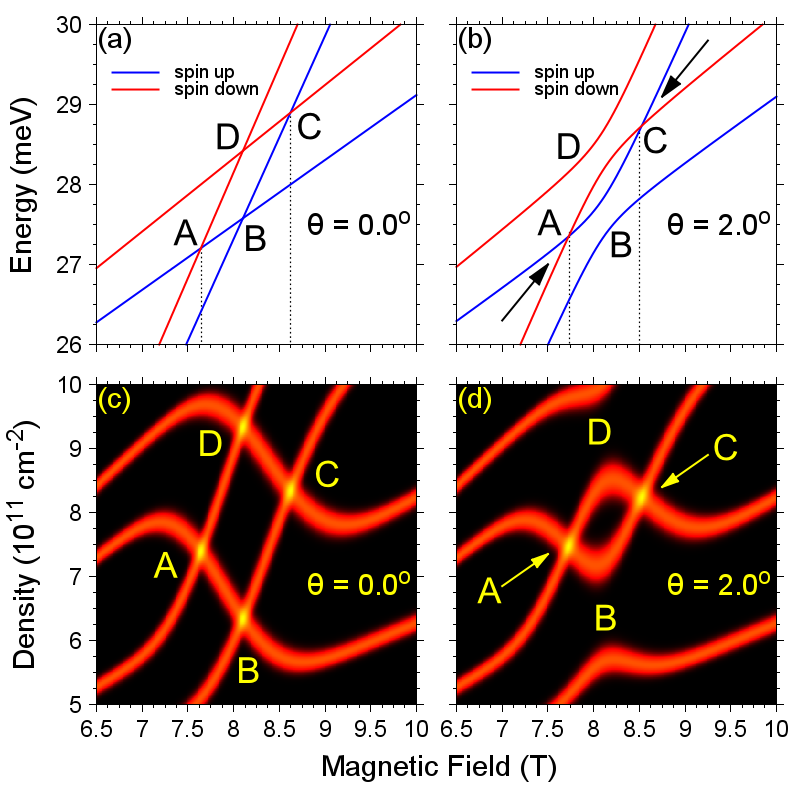}
   \caption{Crossings of non-interacting spin split Landau levels from two subbands for (a) $\theta = 0^\circ$ and (b) $\theta = 2^\circ$. The coupling between LLs due to the tilt angle $\theta$ is introduced via perturbation theory leading to the anticrossings in (b). The dotted lines mark the projections of the LL crossings onto the magnetic field axis. These crossings are mapped into the density--magnetic-field $n_{2D}-B$ diagram of density of states (DOS) as ringlike structures in (c) and (d). In the text we argue that $\rho_{xx}$ and the DOS displays the same topology in the $n_{2D}-B$ diagram. As $\theta$ is increased, the anticrossings shrinks the ring.}
   \label{fig:landaurings}
\end{figure}

For two subband systems, the crossings of the two sets of spin-split LLs from distinct subbands lead to ringlike structures in the density--magnetic-field ($n_{2D}-B$) diagram of the longitudinal resistivity $\rho_{xx}$ \cite{Jaroszynski02, Ritchie96, Muraki01, Zhang05, Ellenberger06, Zhang06_PRB, Zhang06_PRL, ZhangSU4-07,Muraki07,Gerson06}. In the integer quantum Hall regime $\rho_{xx}$ presents peaks (Shubnikov-de Haas oscillations), similar to the density of states (DOS), whenever LLs cross the Fermi level $\varepsilon_F$. As we argue later on, the $n_{2D}-B$ diagram of $\rho_{xx}$ has the same topology as the DOS at $\varepsilon_F$, see the ABCD loops in Figs.~\ref{fig:landaurings}(a)-(d). Interestingly, a recent experimental report \cite{Zhang06_PRB} found that at low enough temperatures the filling factor $\nu = 4$ ring breaks up. This feature was interpreted as a manifestation of quantum Hall ferromagnetic phase transitions in the  system. Moreover, the experimentalists find that a tilted magnetic field with respect to the 2DEG plane makes the ring shrink and fully collapse for large enough angles \cite{ZhangSU4-07,Muraki07}.

Here, we address the collapse of the $\nu=4$ ringlike structure with increasing tilt angle $\theta$ using a non-interacting model, with effective parameters chosen to fit the $\theta = 0^\circ$ data from Ref.~\onlinecite{ZhangSU4-07}. We first obtain the electronic structure by using perturbation theory to treat the LL coupling induced by the tilted magnetic field $B$. We then investigate the collapse by observing the crossing points $A$ and $C$ of the Landau fan diagram for several tilt angles. Our results agree qualitatively with the experimental data in Ref.~\onlinecite{ZhangSU4-07}, indicating that the ring collapse is mainly a single-particle effect due to the coupling of LLs induced by the tilted $B$ field.

The topology of the ringlike structures is determined here through the $n_{2D}-B$ diagram of the DOS. As we argue in the following, the DOS has the same topology of $\rho_{xx}$ in the $n_{2D}-B$ diagram. First we note that the Hall conductivity $\sigma_{xy}$ is  always nonzero, with well defined values at the plateaux $\sigma_{xy} = \nu e^2/h$. In addition, in the linear response regime (Kubo formalism), the longitudinal conductivity $\sigma_{xx}$ has peaks whenever the Fermi level lies within the extended state region of a broadened LL. Since the extended states are at the center of each LL, $\sigma_{xx}$ has the same topology as the DOS. Inverting the conductivity tensor, we obtain $\rho_{xx} = \sigma_{xx}/(\sigma_{xx}^2+\sigma_{xy}^2)$, which has also the topology of the DOS.

We consider a 2DEG formed in a $240\unit{\AA}$ square GaAs quantum well with Al$_{0.3}$Ga$_{0.7}$As barriers, similar to that in Ref.~\onlinecite{ZhangSU4-07}. We assume an effective parabolic confinement for the growth direction to obtain semi-analytical results and emphasize the non-interacting nature of the collapse. The effective mass for GaAs is $m = 0.067m_0$. The measured electron mobility is $\mu_e = 4.1\times 10^5\unit{cm}^2/\unit{Vs}$. The subband energy splitting $\Delta_{SAS} \approx 14\unit{meV}$ and effective $g$-factor $g^* \approx -1.8$ are chosen to adjust the position and size of the ring in the density--magnetic-field diagram\cite{Gerson06}.

The magnetic field $\mathbf{B} = B\sin\theta\hat{y}+B\cos\theta\hat{z}$ is included in our Hamiltonian by using the Landau gauge $\mathbf{A} = zB\sin\theta\hat{x}+xB\cos\theta\hat{y}$,
\begin{equation}
 H = \dfrac{(\mathbf{P}-e\mathbf{A})^2}{2m} + \dfrac{1}{2}m\omega_z^2 z^2 + \dfrac{1}{2}g^*\mu_B B \sigma_z,
\end{equation}
where $\omega_z = \Delta_{SAS}/\hbar$ defines the subband energies, $\mu_B$ is the Bohr magneton and $\sigma_z = \pm 1$. We can rewrite the Hamiltonian as $H = H_z + H_{xy} + \delta H_\theta$, with
\begin{equation}
 \begin{array}{r c l}
 H_z &=& \dfrac{P_z^2}{2m} + \dfrac{1}{2}m(\omega_z^2+\omega_p^2) z^2 + \dfrac{1}{2}g^*\mu_B B \sigma_z,\\
 \\
 H_{xy} &=& \dfrac{P_x^2}{2m} + \dfrac{1}{2}m\omega_c^2(x-\ell_0^2k_y)^2,\\
 \\
 \delta H_\theta &=& \omega_p zP_x,
\end{array}
\end{equation}
in which $\omega_c = eB\cos\theta/m$ is the cyclotron frequency, $\omega_p = eB\sin\theta/m$ the coupling between the planar ($xy$) and growth ($z$) directions due to the tilted magnetic field, $\ell_0 = \sqrt{\hbar/m\omega_c}$ the magnetic length and $k_y$ the wave vector in the $\hat{y}$ direction.
\begin{figure}
   \centering
   \includegraphics[width=8cm,keepaspectratio=true]{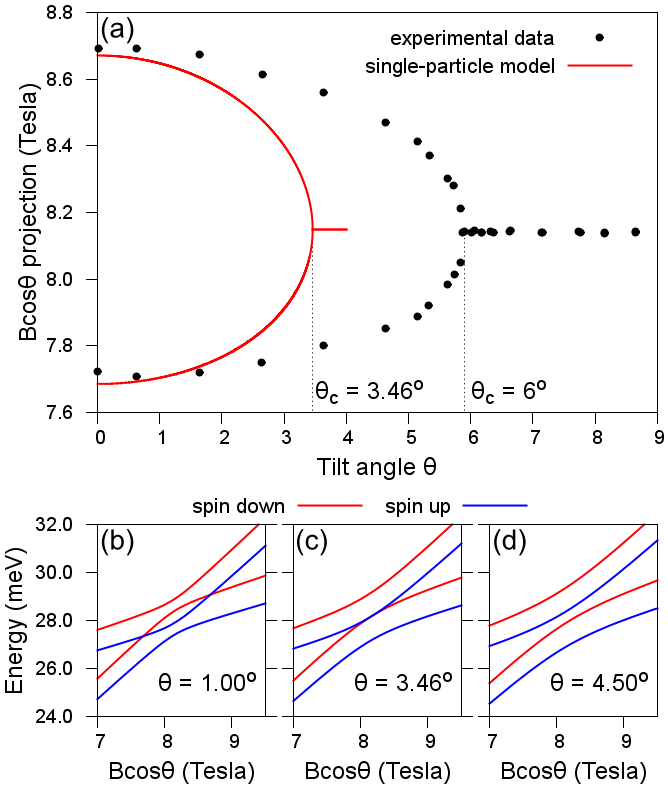}
   \caption{(a) Projections of the LL crossings onto the $B\cos\theta$ axis plotted \textit{versus} the tilt angle $\theta$ illustrating the ring collapse. The black dots are taken from Ref.~\onlinecite{ZhangSU4-07}. The solid line is obtained from the LL crossings in Fig.~\ref{fig:landaurings}. The LL coupling due to the tilted magnetic field is treated via perturbation theory. The calculated collapsing angle $\theta_c = 3.46^\circ$ is about a factor of two smaller than the experimental one ($\approx 6^\circ$); however, both exhibit similar features. Panels (b)-(d): evolution of the LL anti-crossings as $\theta$ is increased.}
   \label{fig:colapso}
\end{figure}

Since $\omega_p \propto \sin\theta$ and $\theta$ is small, $\delta H_\theta$ can be straightforwardly accounted for via perturbation theory in the basis set $\{\ket{jnk_y\sigma_z}\}$ of the non-perturbed Hamiltonian $H_0 = H_z + H_{xy}$. Note that
\begin{equation}
\begin{array}{r c l}
   \braket{\mathbf{r}}{jnk_y\sigma_z} &=& \dfrac{e^{ik_yy}}{\sqrt{L_y}}\chi_n\left(x-\ell_0 k_y\right)\varphi_j(z)\ket{\sigma_z},
\end{array}
\end{equation}
where $\chi_n(x)$ and $\varphi_j(z)$ are solutions of the harmonic oscillator Hamiltonians in the $x$ and $z$ directions, respectively. Here, $j$ and $n$ are the subband and Landau level indices, respectively. Defining $\omega_\theta = \sqrt{\omega_z^2 + \omega_p^2}$ for simplicity, we find that the zeroth-order energies are
\begin{equation}
 \varepsilon_{jn\sigma_z}^{(0)} = \left(j+\dfrac{1}{2}\right)\hbar\omega_\theta + \left(n+\dfrac{1}{2}\right)\hbar\omega_c + \dfrac{1}{2}g^*\mu_B B \sigma_z.
\end{equation}
Since the energy is degenerate in $k_y$ and the perturbation does not couple $k_y$, we will omit this index from now on.

The perturbation $\delta H_\theta$ only couples consecutive subbands ($j^\prime = j\pm 1$) and LLs ($n^\prime = n \pm 1$), hence
\begin{equation}
\begin{array}{r c l}
 \bra{j^\prime n^\prime\sigma_z} &\omega_p zP_x& \ket{jn\sigma_z} = i \dfrac{\hbar\omega_p}{2}\dfrac{\ell_z}{\ell_0}\times\\
 \\
 & & \left[ \sqrt{j+1}\delta_{j^\prime,j+1}+\sqrt{j}\delta_{j^\prime,j-1}\right] \times\\
 \\
 & & \left[ \sqrt{n+1}\delta_{n^\prime,n+1}-\sqrt{n}\delta_{n^\prime,n-1}\right],
\end{array}
\end{equation}
where $\ell_z = \sqrt{\hbar/m\omega_\theta}$. Thus, the first-order corrections near the crossings of $\varepsilon_{jn\sigma_z}^{(0)}$ and $\varepsilon_{j^\prime n^\prime\sigma_z}^{(0)}$ are\cite{Bastard88}

\begin{equation}
\begin{array}{r c l}
 \varepsilon^{(1)}_{\pm\sigma_z} &=& \dfrac{\varepsilon_{jn\sigma_z}^{(0)}+\varepsilon_{j^\prime n^\prime\sigma_z}^{(0)}}{2} \pm \\
 \\
 &\pm& \sqrt{ \left(\dfrac{\varepsilon_{jn\sigma_z}^{(0)}-\varepsilon_{j^\prime n^\prime\sigma_z}^{(0)}}{2}\right)^2 + \Delta^2},
\end{array}
\label{eq:energy2}
\end{equation}
where $\Delta^2 = \left|\bra{j^\prime n^\prime\sigma_z} \omega_p zP_x \ket{jn\sigma_z}\right|^2$.

For simplicity, we assume the DOS to be a set of broadened gaussians, due to impurity scattering \cite{Gerhardts76,Ando83,AndoReview}, centered at the LL energies $\varepsilon^{(1)}_{\pm\sigma_z}$, i.e.,

\begin{equation}
 g(\varepsilon_F) = \dfrac{eB}{h} \sum_{\pm,\sigma_z} \dfrac{\exp{\left[-\dfrac{(\varepsilon_F-\varepsilon^{(1)}_{\pm\sigma_z})^2}{2\Gamma^2}\right]}}{\sqrt{2\pi\Gamma^2}},
\end{equation}
where $\Gamma$ is the half-width of the gaussian broadening. In the short-range scattering approximation\cite{Ando83} the LL broadening is related to the electron mobility $\mu_e$ as $\Gamma \rightarrow \Gamma_{\mu_e} \propto \sqrt{B/\mu_e} = 0.210\sqrt{B}\unit{meV}$ ($B$ in Tesla). In transport measurements however, only the extended states near the center of each LL contribute to $\rho_{xx}$
\footnote{We consider that all electrons are extended (conducting) [see, R.~E.~Prange, Phys.~Rev.~B \textbf{23}, R4802 (1981)].}. We model these regions of extended states as a DOS with smaller broadening\cite{Freire07PRL} $\Gamma \rightarrow \Gamma_{ext} = 0.110\unit{meV}$.

Figures \ref{fig:landaurings}(a)-(b) show Landau fan diagrams for $\theta = 0^\circ$ and $\theta = 2^\circ$ calculated using Eq.~(\ref{eq:energy2}). For increasing tilt angles, the anticrossings induced by the inter LL couplings also  increase thus making the A and C crossings in Fig. 1(d) move closer together. Figures 1(c) and 1(d) show the $n_{2D}-B$ diagram of $g(\varepsilon_F)$. We can see that the ring in 1(d) shrinks as the anticrossings in 1(b) increase for larger tilt angles.

Figure 2(a) shows a plot of the projection of the points A and C in Fig.~\ref{fig:landaurings}, on the magnetic field axis, \textit{versus} $\theta$. The solid line is obtained from our non-interacting model and the black circles are taken from the experimental data of Ref.~\onlinecite{ZhangSU4-07}. The collapsing angle in our calculation, $\theta_c \approx 3.46^\circ$ [Fig.~\ref{fig:colapso}(c)] is smaller than the experimental one $\theta_c \approx 6^\circ$. Despite this discrepancy, our results, solid line in Fig.~2(a), have similar behavior to the experimental data. Particularly, near the collapsing angle $\theta_c$, both show sharp transitions.

Note that as the angle $\theta$ is increased, opposite spin LLs are brought close together, Figs.~\ref{fig:colapso}(b)-(d). 
This is a configuration in which the interplay of the temperature, Coulomb and Zeeman energy scales is most important. Hence, deviations from our results due to exchange-correlation effects are expected \footnote{G.~J.~Ferreira, H.~J.~P.~Freire, and J.~C.~Egues, Phys.~Rev.~Lett. \textbf{104}, 066803 (2010).}.

In summary, we have investigated the collapse of the $\nu = 4$ ring using a non-interacting model which perturbatively accounts for the inter LL coupling due to the tilted $B$ field. Our results show that the anticrossings induced by this coupling make the ring shrink for increasing tilt angles. Our calculated full collapse angle $\theta_c = 3.46^\circ$ for the $\nu=4$ ring is about half of the experimental one $\theta_c = 6^\circ$ in Ref.~\onlinecite{ZhangSU4-07}. However, we find that the overall description of the ring shrink and collapse is essentially a single-particle effect due to the inter LL coupling induced by the tilted magnetic field.

GJF acknowledges useful conversations with H.~J.~P.~Freire, X.~C.~Zhang and T.~Ihn. This project is supported by Fapesp and CNPq (Brazilian funding agencies).


\end{document}